\documentclass[prd,11pt]{revtex4}
\newcommand{\ww}{\mbox{\tiny $\wedge$}}

\usepackage{epsfig}

\begin{document}

\title{Charged Black Holes in Gauss-Bonnet Extended Gravity.}
\author{M\'aximo Ba\~nados}
\affiliation{Departamento de F{\'\i}sica, P. Universidad Cat\'olica de Chile, \\
Casilla 306, Santiago 22,Chile. \\ {\tt mbanados@fis.puc.cl}    }

\begin{abstract}

Charged Black holes in Gauss-Bonnet extended gravity are studied.  The electromagnetic
field is coupled non-minimally, as in $U(2,2)$ Chern-Simons theory. We find that the
geometrical properties of the solution exhibit ``phase transitions" as one varies the
mass and charge. The full phase diagram for all values of the ADM mass and charge is
displayed.

%\pacs{04.50.+h, 04.65.+e}

\end{abstract}

\maketitle

%------------------------------------
\section{Introduction, Main Results and Conclusions}

In a five dimensional Universe, the Gauss-Bonnet density $\:\sqrt{-g}( R^2 - 4
R^{\mu\nu}R_{\mu\nu} + R^{\mu\nu\lambda\rho}R_{\mu\nu\lambda\rho}) \:$ cannot be omitted
in the gravitational Lagrangian.  This term is covariant, its associated Einstein tensor
is conserved, and, despite being quadratic in the curvature tensor, yields second order
field equations for the metric \cite{Lovelock,Zwiebach}. The most general action for
Gravity in five dimensions is then
\begin{equation}\label{I}
    I[g_{\mu\nu}] = \int_{M_5} \sqrt{-g} \left[  \alpha_0  + \alpha_1 \:R  +
    \alpha_2( R^2 - 4 R^{\mu\nu}R_{\mu\nu} +
    R^{\mu\nu\lambda\rho}R_{\mu\nu\lambda\rho}) \right]
\end{equation}
The presence of this term of course changes the dynamical equations, and many aspects of
general relativity have to revisited. This issue is particularly relevant in the context
of brane worlds models, and many papers have recently been devoted to the subject
\cite{Branes}.

The simplest problem that can be analyzed in a closed form is the spherically symmetric
five-dimensional black hole spacetime,
\begin{equation}
    ds^2 = - N^2 dt^2 + {dr^2 \over f^2} + r^2 d\Omega_3.\label{BD}
\end{equation}
Although the equations can be solved for arbitrary values of the three couplings
\cite{Boulware-D,Wheeler,BTZ2}, we are interested in the rol of the Gauss-Bonnet term and
then we set, for simplicity, $\alpha_1 =1$, $\alpha_0  = 0$ and $\alpha_2>0$. The
equations of motion yield \cite{Boulware-D,Other} \footnote{The coefficient $\alpha_2$
appearing here differs from that in (\ref{I}) by a numerical constant.   Note that
$\alpha_1=1$ implies, in five dimensions, $[mass]=[length^2]=[\alpha_2]$ }
\begin{equation}\label{N}
    N^2 = f^2 = 1+ {1 \over \alpha_2}\left( r^2 - \sqrt{ r^4 +  4 \alpha_2 M}\right)
\end{equation}
where $M$ is an integration constant, that will be seen to be the ADM mass.

We first note that for $M=0$ the metric reduces to flat space, which is the stable
background for this theory \cite{Boulware-D}. It also follows that for $r^4 \gg 4M
\alpha_2$,
\begin{equation}\label{NSch}
    N^2 \simeq 1- {2 M \over r^2}
\end{equation}
showing that, asymptotically, (\ref{N}) approaches the five dimensional Schwarzschild
metric. Incidentally, note that if $ M \gg \alpha_2$ then $r^4 \gg 4M \alpha_2$ would
hold all the way to the Schwarzschild horizon at $r_+=\sqrt{2M}$. If $\alpha_2=0$,
(\ref{NSch}) becomes an exact solution.

The function (\ref{N}) has some interesting properties. The value of $N^2$  at the
singularity $r=0$ is finite,
$$
N^2(0) = 1 - \sqrt{{4M \over \alpha_2}},
$$
although the curvature is still singular\footnote{While for the cone metric $ ds^2 =
\alpha\; dr^2 + r^2 d\phi^2$ the curvature is concentrated at $r=0$, for the sphere $ds^2
= \alpha\: dr^2 + r^2( d\theta^2 + \sin^2\theta\: d\phi^2)$, the scalar curvature is $R =
(\alpha-1) / (\alpha r^2)$.}. The location of the horizon, $N^2(r_+)=0$, is
\begin{equation}\label{r+}
    r_+ = \sqrt{ 2M - {\alpha_2 \over 2} }.
\end{equation}
We see from this expression that the horizon exists only for $M>\alpha_2/4$. We thus find
a mass gap separating flat space from the spectrum of black holes:
\begin{equation}\label{gap0}
 \begin{array}{rll}
     M=0 & ~ & \mbox{Flat space} \\
      0<M\leq\alpha_2/4 & ~& \mbox{Naked Singularities} \\
      M > \alpha_2/4 &~ &\mbox{Black holes} \\
    \end{array}
\end{equation}
The mass gap appears in all odd dimensional theories containing the highest Lovelock
\cite{Lovelock} term. In three dimensions, this term is just the Hilbert term and the
mass gap is present \cite{BTZ}. In this case, however, the ``naked singularities" have a
sensible interpretation in terms of particles \cite{Deser-JtH}.

In this paper we add electric charge $q$ to this black hole and study the corresponding
spectrum.  We shall see that the solution has some peculiarities not present in usual
charged black holes.

\begin{figure}[ht]
\centerline{\epsfig{file=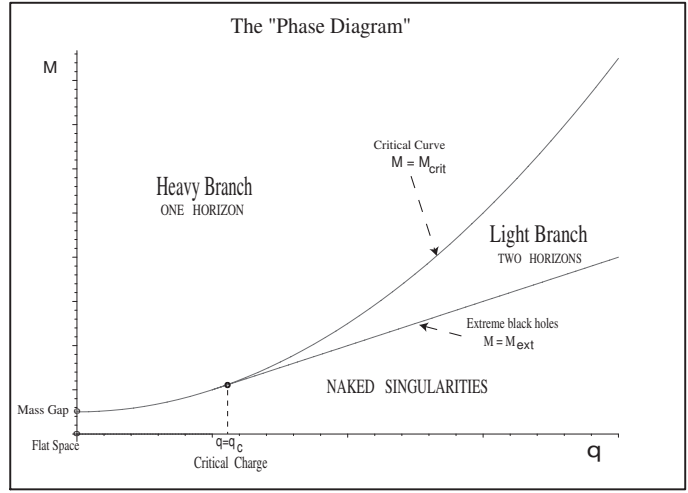,width=9cm,angle=0}} \caption{The phase diagram.
For each value of $q>q_c$ there exists two black holes phases. Flat space is located at
$q=M=0$, and it is disconnected from the black hole spectrum by a set of naked
singularities. } \label{Phases}
\end{figure}

As a first surprise, the addition of charge does not imply the existence of two horizons.
There are open regions in the plane $\{M,q\}$ having non-extremal black holes with {\it
one} horizon. In order to find solutions with two horizons, $q$ has to be bigger than a
certain critical value, $q>q_c$.  Fig. \ref{Phases} gives a summary of the properties of
various solutions obtained by varying the values of $M$ and $q$.

In one region of the space of parameters, we find a set of non-extremal charged black
holes having only one horizon.  We call this region the ``Heavy Branch" because it is
defined by the condition  $M > M_{crit}$.  For masses within the range
\begin{equation}\label{M}
    M_{crit} \geq  M > M_{ext}, \ \ \ \ \ \ \ \ \ \  q>q_c
\end{equation}
we find the ``Light Branch" with black holes with two horizons. At $M=M_{ext}$, we find
extremal black holes with only one horizon (and zero Hawking temperature). Below the
extreme value, the solution represents a naked singularity.  There is also a ``critical
charge" below which the light branch ceases to exist. Flat space is located at $M=q=0$,
and thus the mass gap persists in the charged solution.

The terms ``phase structure" and ``phase transitions" are used here only in analogy with
the statistical mechanics concept, without implying a direct connection. Of course, given
the thermodynamical properties of black holes, this may turn out to be more than an
analogy but we shall not study this issue here.

To avoid future confusions, we stress that the action considered in this paper is not the
usual minimally coupled Einstein-Maxwell system (plus a Gauss-Bonnet term). Those
solutions were studied in \cite{BTZ2} and do not exhibit this phase structure. Instead,
we consider a five-dimensional Chern-Simons theory for the group $U(2,2)$
\cite{Chamseddine}, which has a sensible interpretation as a gravitational plus
electromagnetism theory, with a Gauss-Bonnet term. This interpretation, however, requires
a symmetry breaking term because otherwise the equations of motion differ from the usual
ones even asymptotically. This point was discussed in detail in \cite{B}.

The application of Chern-Simons theories to gravity has been discussed several times in
the literature and we shall not repeat it here. The first constructions were reported in
three dimensions in \cite{Achucarro}, and the same idea was then applied in
\cite{Chamseddine} to five dimensions. See \cite{CS} for other aspects.

For the purposes of this paper, we refer the reader to Ref. \cite{B} were many details
omitted here can be found. In particular, the asymptotic form of the charged black holes
was already reported in that Reference. The goal of the present paper is to display the
{\it exact} solution for an arbitrary mass $M$ and electric charge $q$, and study the
associated phase space.

%-----------------------------------------------
\section{The equations and their solution}
\label{Charged}

\subsection{The equations and spherically symmetric ansatz}

We start by writing down the equations of motion associated to a Chern-Simons theory for
the group $U(2,2)$. This group contains $SO(4,2)$ which would give pure gravity with a
Gauss-Bonnet term. The extension to $U(2,2)$ incorporates an Abelian one form, that we
interpreted as an electomagnetic field, coupled non-minimally to gravity. See \cite{B}
for a detailed analysis, and other motivations to study this system.

A key issue in the analysis (and discussed in \cite{B}) is the fact that the Chern-Simons
equations of motion do not provide a sensible theory for the $SU(2,2)$ field. For
example, there is no linearized theory and the metric is not asymptotically
Schwarzschild.   Given the strong topological roots of the Chern-Simons construction, it
is still attractive as a field theory, and one would like to know if the equations can be
``repaired" by some mechanism, hopefully within the same theory.  Some progress in this
direction was reported in \cite{Alfaro-B}.

Here we follow \cite{B} in which a symmetry breaking term is added to the action. The
Chern-Simons equations then becomes closer to the real world and one can start asking
questions such as what is the structure of black holes, and what is the nature of the
couplings between the gravitational and internal $U(N)$ gauge fields degrees of freedom.
In this paper we concentrate on the coupling between the gravitational and $U(1)$ field.

The symmetry breaking term added in \cite{B} consists in a cosmological term (vacuum
energy) and it is parameterised by a real number $\tau$. Let $e^a$ be the five
dimensional vielbein one-form, $w^{ab}$ the spin connection, and $A$ the Abelian
one-form. The equations of motion following from the $U(2,2)$ Chern-Simons theory
(including the symmetry breaking term proportional to $\tau$) are,
\begin{eqnarray}
\epsilon_{abcde} [ (R^{ab}+e^ae^b) \mbox{\tiny $\wedge$} ( R^{cd}+e^ae^b)- \tau^2 e^a\ww
e^b\ww e^c \ww e^d ]  &=& - 4 \,  T_e \mbox{\tiny $\wedge$} F
\label{e} \\
\epsilon_{abcde} ( R^{ab}+e^ae^b) \mbox{\tiny $\wedge$} T^c &=& ( R_{de}+e_de_e)
\mbox{\tiny $\wedge$} F
\label{w} \\
{1 \over 2} R^{ab} \mbox{\tiny $\wedge$} R_{ab} - d(e_a \mbox{\tiny $\wedge$} T^a) &=& F
\mbox{\tiny $\wedge$} F \label{A}
\end{eqnarray}
where $R^{ab}=dw^{ab} + w^a_{\ c} w^{cb}, T^a = de^a + w^a_{\ b} e^b$ and $F=dA$. For
$\tau=0$, these equations are equivalent to ${\cal F} \wedge {\cal F}=0$ with ${\cal F}
\in U(2,2)$ which are the exact Chern-Simons equations.

Since we are interested  in black hole solutions, we write the ansatz for the metric and
gauge field with spherical symmetry
\begin{eqnarray}
ds^2 &=& - N(r)^2 dt^2 + f^2(r) dr^2  + r^2 d\Omega_3, \label{metric2} \\
   A &=& -\phi(r) dt,
\end{eqnarray}
where $N,f$ and $\phi$ are functions to be determined.

As shown in \cite{B}, the dynamics of the Abelian form $A$ is linked to the torsion
tensor. If we assume from the very beginning that $T^a=0$, then the equations of motion
for $A$ do not give Maxwell's equations in any limit. On the contrary, letting $T^a$ to
be different from zero yields a system of equations that can be analyzed perturbatively
and one obtains, to first order, Maxwell's theory for the potential $\phi(r)$. The
relationship between the torsion $T^a$ and the Maxwell field $F$ is encoded in Eq.
(\ref{w}). To first order we ignore the right hand side of (\ref{e}) and find the
gravitational background (AdS space) $R^{ab}= (\tau-1)e^ae^b$. Replacing in (\ref{w}) it
implies \cite{B}
\begin{equation}\label{dual}
    e_a \wedge T^a =~ ^* \hspace{0cm} F,
\end{equation}
where $^*$ represents Hodge's dual. Replacing (\ref{dual}) into (\ref{A})  one obtains
the usual five-dimensional Maxwell-Chern-Simons system $d ^* \!\! F = F\wedge F$.   This
is, in short, the mechanism that transforms the ``topological" one-form $A$ into a
radiating Maxwell field.

Incorporating the back reaction from the right hand side of (\ref{e}) generates
corrections to (\ref{dual}), and the Maxwell equation. It is precisely the role of these
corrections what we aim to investigate in this paper.

Let us then assume that the torsion is not zero, and let $ w^{ab} = w^{ab}(e) +
\kappa^{ab}$ where $w^{ab}(e)$ is the solution to the equation $de^a + w^a_{\ b}(e) \ww
e^b=0$, and only depends on the metric.  $\kappa$ parameterizes a non zero torsion by
$T^a=\kappa^a_{\ b} \ww e^b$. In order to prescribe a spherically symmetric ansatz for
$\kappa^{ab}$, it is convenient to express all indices in the coordinate basis $
\kappa_{\alpha\beta\ \mu } = e_{a\alpha}\, \kappa^{ab}_{\ \ \mu} \, e_{b\beta}$. The
correct ansatz with spherical symmetry for this tensor follows by studying the equations
of motion order by order starting from the AdS vacuum. The details were given in
\cite{B}, and the result is,
\begin{equation}
\kappa_{\mu\nu\; \lambda } = {\psi(r) \over 2 \sqrt{h} }\; \epsilon_{\mu\nu\lambda}^{\ \
\ \ \, \alpha\beta} \; z_{\alpha\beta} - z_{\mu\nu} U_\lambda + 2g_{\lambda[\mu} V_{\nu]}
\end{equation}
where $\epsilon^{\mu\nu\lambda\alpha\beta}$ is the Levi-Civita tensor with
$\epsilon^{tr\theta_1\theta_2\theta_3}=1$, $\sqrt{h} = r^3 \sin^2\theta_1\sin\theta_2 $,
and $z =dt \ww dr,U = \beta(r) \, dt $, $V = \alpha(r) \, dr$. The torsion then
contributes with three more functions of $r$ to be determined by the equations of motion,
namely,  $\alpha(r),\beta(r)$ and $\psi(r)$. $\psi$ is closely related to the electric
field (see Eq. (\ref{dual})), while $\alpha$ and $\beta$ are auxiliary fields which will
be eliminated algebraically from their own equations of motion.

We shall see that this ansatz does provide an exact solution to the full $U(2,2)$ system.

%-----------------------------------------
\subsection{The spherically symmetric reduced equations}

In this section we present some of the details in finding the equations and their
solution. The reader interested only the final result can jump to the next section where
the final form of the metric is displayed and its properties analyzed.

It is a direct but long calculation to replace in the equations of motion
(\ref{e},\ref{A}) the ansatz with spherical symmetry shown in the last paragraph. The
resulting equations turns out to be extremely complicated.  Some simplification can be
achieved by making field redefinitions which simplify the expressions for the curvature
$R^{ab}$ and torsion $T^a$. These field redefinitions  involve all variables. We
transform $ \{\alpha,\beta,\ N,f,\ \psi,\phi,\} \rightarrow \{\alpha_1,\beta_1, N_2,f_2,\
\psi_1,\phi_1,\ \}, $ according to
\begin{eqnarray}
\alpha &=& {\alpha_1 - 1 \over r} \nonumber\\
\beta &=& N \left( {fN\beta_1 \over g} + N' + N\alpha  \right)\nonumber\\
f &=& {\alpha_1 \over f_1},  \hspace{2cm} f_1 = \sqrt{1+r^2 - \psi_1^2 + f_2}  \nonumber\\
N &=&   f_1 \; N_1,   \hspace{1.5cm} N_1  = e^{\int^r N_2} \nonumber\\
\phi &=& 4 \int^r dr\; ( N_1 \psi_1\Phi_1 ) \label{redef}  \\
\psi &=& {\psi_1 \over r} \nonumber
\end{eqnarray}
Note that the horizon structure will be controlled by the zero'es of the function $f_1$.

Inserting the above ansatz into the equations of motion, we find two sets of equations
which can be analyzed and solved separately.

%-----------------------------------
\subsubsection{The $\alpha_1,\psi_1,f_2$ system}

The first set of equations involves only the functions $\{\alpha_1,\psi_1,f_2\}$. The
equations are the following (prime indicates radial derivative):
\begin{eqnarray}
  (4\psi_1^3 - 4r^2 \psi_1 - 4\psi_1 - 2f_2 \psi_1) \psi_1' - 2f_2 r + 2r f_2 \alpha_1 +
   2\tau^2r^3 \alpha_1  &=& 0 \label{H} \\
  \alpha_1 f_2 - f_2 -2r \psi_1 \psi_1' &=& 0 \label{e2} \\
  2\psi_1^2 \psi_1' - (\psi_1 f_2)' &=& 0 \label{G}
\end{eqnarray}
We note that (\ref{e2}) and (\ref{G}) can be easily solved. We find, respectively,
\begin{eqnarray}
  \alpha_1 &=& 1 +  {2r\psi_1\psi_1' \over f_2} \label{alpha1}\\
  f_2 &=& {2 \over 3} \psi_1^2 + {q_0 \over \psi_1} \label{f2}
\end{eqnarray}
where $q_0$ is an integration constant that will be related to the electric charge. It
will be convenient in what follows to do yet one more redefinition,
\begin{equation}\label{Psi}
    \psi_1(r) = -{q_0 \over \ {\Psi}(r)}, \ \ \ \    q_0 = {\sqrt{6}\: q \over 2}
\end{equation}
where ${\Psi}(r)$ is a new function of $r$. Replacing (\ref{f2}) into (\ref{alpha1}) we
find the following expression for $\alpha_1$,
\begin{equation}\label{alpha1i}
    \alpha_1 = {(P \, r)' \over P}
\end{equation}
where $P$ is a short hand for
\begin{equation}\label{P}
    P = {\Psi^3 -q^2 \over {\Psi}^3}.
\end{equation}

Finally, we replace (\ref{alpha1}) and (\ref{f2}) in (\ref{H}) and obtain a closed
equation for the function ${\Psi}$:
\begin{equation}\label{Hi}
    \left( 6\,{q}^{4}  {\Psi}^{2}+3\,{q}^{2}
    {\Psi} ^{6}-6\,{q}^{2} {r}^{4} {\Psi}^{4}-6\,{q}^{2}
 {\Psi}^{5}-4\,{q}^{6}+ {\Psi}^{9} \right) {\frac {d{\Psi}}{dr}}  +
 2\tau^2 \,r^3 {\Psi}^5 ({q}^{2}  -{\Psi}^{3}) =0
\end{equation}
This last equation is linear in $\Psi'$. The integral can be done explicitly and we find
the implicit solution for $\Psi(r)$,
\begin{eqnarray}
{\tau^2 \over 2}\,\left({\Psi^3 -q^2 \over {\Psi}^3}\right)^4 r^4 -{1 \over 2}\, {\Psi}
^{2} - 3\,{\frac {{q}^{2}}{ \ {\Psi} ^{2}}}+ {4 \over 7}\,{\frac {{q}^{6}}{ \ {\Psi}
^{7}}}-{9 \over 2}\,{\frac {{q}^{6}}{ \ {\Psi} ^{8}}}+{\frac {9}{10}}\,{\frac {{q}^{8}}{
 {\Psi} ^{10}}}+{\frac {24}{11}}
\,{\frac {{q}^{8}}{ \ {\Psi} ^{11}} } + && \nonumber\\ +{\frac {24}{5}}\,{\frac
{{q}^{4}}{
 {\Psi} ^{5}}}-{\frac {12}{13}}\,{\frac {{q}^{10}}{ {\Psi}^{13}}}+ {1 \over 4}\,{\frac
{{q}^{12}}{  {\Psi} ^{16}}}-{3 \over 2}\,{\frac {{q}^ {4}}{ {\Psi} ^{4}}}-{3 \over
7}\,{\frac { {q}^{10}}{ {\Psi} ^{14}}} &=& -2 M \label{Eqr}
\end{eqnarray}
where $M$ is an integration constant that will be seen to be the ADM mass of the
solution. This is an algebraic equation that should be inverted to find $\Psi(r)$.

%-----------------------------------
\subsubsection{The $\beta_1,N_2,\Phi$ system }

We now proceed to find expressions for $\beta_1,N_2$ and $\Phi_1$ in terms of ${\Psi}$.
The three remaining equations are:
\begin{eqnarray}
 0 &=&  {\it f_2}  r + \beta_1 f_2  + {r}^{3}
 \label{1} \\
 0 &=& {\it \alpha_1}  {\it f_2}  + {\it \beta_1'} {\it f_2}  + {\it f_2'}
 { \it \beta_1}  + {\it N_2}  {\it \beta_1}
  {\it f_2}  -2\,{\it \psi_1}   {\it \psi_1'}   {
\it \beta_1} +  \nonumber \\& \ &  +2\,{r}^{2}+4\,r  {\it \psi_1} ^{2}{\it \Phi_1}
-2\,r{\it \psi_1}  {\it \psi_1'}  -2\,{\it \alpha_1}  {\it \beta_1}  r+ r f_2' +{\it
\alpha_1} {r}^{2}+2\, r{\it \beta_1}
 \label{2} \\
0 &=&  2\,{\it \alpha_1}  {\it \beta_1}  -2\,{ \it f_2}  {\it \Phi_1}  -2\,{\it N_2}
   {\it \psi_1} ^{2}+2
\,{\it N_2}  +{\it f_2'}+    \nonumber \\&\ & + 2\,{\it N_2}  {\it f_2} -2\,{\it \psi_1}
  {\it \psi_1'}  + 2\,r{\it \beta_1'}  +2\,r{\it \alpha_1}
  + 2\,r+2\,r{\it N_2}  {\it \beta_1}
+2\,{r}^{2}{\it N_2}
 \label{3}
\end{eqnarray}

This set of equations can also be solved in a closed form.  We shall not go into the
details on how to find the solution, we only quote the result. Using the value of
$\alpha_1$ found in the previous paragraph, we find that Eqns. (\ref{1}-\ref{3}) are
solved by:
\begin{eqnarray}
  \beta_1 &=& - {r(f_2 + r^2) \over f_2} \\
  N_2  &=& {8 \psi_1 \psi_1' \over f_2} \\
  \Phi_1 &=& {1 \over 2}{(3f_2^2 -2r^4) \psi_1' \over f_2^2 \psi_1}
\end{eqnarray}

Note that making the redefinition (\ref{Psi}), and using the solution (\ref{f2}) for
$f_2$ we have
$$ N_1 = e^{\int N_2} = \left({ \Psi^3 - q^2 \over \Psi^3}\right)^4 $$
This completely solve the problem. All functions are known in terms of $r$ and $\Psi$,
and $\Psi$ is known in terms of $r$ by (\ref{Eqr}).

%----------------------------------------
\section{Charged black holes and a phase transition}
\label{Analysis}

\subsection{The metric}

Let us summarize the results of the analysis of the equations of motion.  The metric
ansatz was
\begin{eqnarray}
ds^2 = - N(r)^2 dt^2 + f^2(r) dr^2  + r^2 d\Omega_3. \label{metric2i}
\end{eqnarray}
The functions $N$ and $f$ are fixed by the equations of motion as (prime denotes radial
derivative)
\begin{eqnarray}
  N^2   &=& P^8\: f_1^2, \nonumber\\
  f^2 &=&  \left[{(P \:r)' \over P}\right]^2 \: {1 \over f_1^2} \label{fh2}
 \end{eqnarray}
where
\begin{eqnarray}
    f_1^2 = 1+ r^2 - {1 \over 2} \: {q^2 \over {\Psi}^2} - {\Psi}, \label{f1i}
\end{eqnarray}
and $P$ is given in (\ref{P}).  Finally, $\Psi$ is a function of $r$ defined by the
algebraic equation (\ref{Eqr}).

In some applications it may be convenient to define a new radial coordinate, as suggested
by Eqns. (\ref{Eqr}) and (\ref{fh2}),
\begin{equation}\label{rho}
    \rho = P\, r
\end{equation}
In terms of this new coordinate the metric takes a simple form,
\begin{equation}\label{metrici}
    ds^2 = - P^8 f_1^2 dt^2 + {1 \over P^2} \left( {d\rho^2 \over f_1^2} + \rho^2 d\Omega_3^2
    \right)
\end{equation}
and the relation (\ref{Eqr}) becomes
\begin{eqnarray}\label{rhoi}
    \tau^2\rho^4 +4M &=& {\Psi}^{2}+6\,{\frac {{q}^{2}}{{\Psi}^{2}}}-{\frac {8}{7}}
\,{\frac {{q}^{6}}{{\Psi}^{7}}}+9\,{\frac {{q}^{6}}{{\Psi} ^{8}}}-{9 \over 5}\,{\frac
{{q}^{8}}{{\Psi}^{10}}}-{\frac {48}{11}}\,{ \frac {{q}^{8}}{{\Psi}^{11}}}-  \nonumber\\
&& {\frac {48}{5}}\,{\frac {{q}^{4}}{ {\Psi}^{5}}}+{\frac {24}{13}}\,{\frac
{{q}^{10}}{{\Psi}^{ 13}}}-{1 \over 2}\,{\frac {{q}^{12}}{{\Psi}^{16}}}+ 3\,{\frac
{{q}^{4}}{ {\Psi}^4} }+{6 \over 7}\,{\frac {{q}^{10}}{{\Psi}^{14}}}
\end{eqnarray}

From now we shall only consider the case $\tau=1$. This is only for simplicity in some
calculations, but it does not affect the main conclusions.

%-------------------------------
\subsection{Known limits of the solution. Reissner-Nordstrom Solution}

Since the exact solution displayed in the previous paragraph is rather complicated, as a
first check we analyze how this solution reduces to the known ones, in various limits. We
first study the uncharged solution, found in \cite{Boulware-D} and discussed in the
introduction. Then we show how in the limit of small charges and large radial coordinate
$r$, we recover the usual Reissner-Nordstrom spacetime.

Consider first the uncharged solution with $q=0$. In this case, $P=1$ (hence $r=\rho$),
we can solve ${\Psi}$ explicitly, ${\Psi} = \sqrt{r^4+4M}$, and obtain for $f_1$ the
closed expression,
\begin{eqnarray}\label{f1ii}
    f_1^2 &=& 1+r^2 - \sqrt{r^4 + 4M },
\end{eqnarray}
representing the uncharged solution \cite{Boulware-D}, described in the introduction,
with $\alpha_2=1$.

Consider now the charge as a small parameter, and seek for a perturbative solution in
$q^2$ to the equation (\ref{Eqr}). Let
\begin{equation}\label{pert}
    {\Psi}(r) = \sqrt{r^4+4M} + q^2\: h_1(r)
\end{equation}
and solve (\ref{Eqr}) keeping only the linear terms in $q^2$.  We find for the first
order perturbation
\begin{equation}\label{h1}
    h_1(r) = -{2r^4+3\sqrt{r^4+4M}\over (r^4+4M)^2}
\end{equation}
Replacing in (\ref{fh2}), and taking the limit $r^4 \gg 4M$, we find
\begin{eqnarray}
  N^2 = {1 \over f^2} =  1 - {2M \over r^2} + {3 \over 2} \: {q^2 \over r^4} +
  {\cal O}\left( {1 \over r^6} \right)
 \end{eqnarray}
coinciding exactly with the Reissner-Nordstrom spacetime. We can also see that the
parameter that we called $q$ is in fact the electric charge, up to a normalization.

%------------------------------------------------
\subsection{The origin, the Kasner singularity, and physical range of radial coordinate}

The metric (\ref{metrici}) has a curvature singularity at the origin $\rho=0$. There is
another singularity  at the point where $P$  vanishes ($P$ was defined in (\ref{P})) ,
\begin{equation}\label{sin}
    {\Psi}^3 - q^2 = 0.
\end{equation}
At this singularity, proper times are shrink to zero while spacelike separations are
stretch to infinity. The volume element, however, remains finite, $ \det g \sim P^8
\times ( P^{-2})^4 =1$. For this reason, we call this point the ``Kasner singularity".

Both singularities can be shown to be physical in the sense that the components of the
curvature tensor in an inertial frame diverge. It is then important to ask whether these
singularities are protected by horizons.

The analysis of existence of horizons is greatly simplified by noticing that the function
$\Psi$ can be used as a radial coordinate. Although the algebraic relation (\ref{rhoi})
between $\rho^4$ and $\Psi$ is quite untractable and attempts to invert it explicitly are
hopeless, we can in fact show that in the domain of interest, it is an invertible
function.

We first note that the derivative of (\ref{rhoi}) can be factorized in the form,
\begin{equation}\label{rhoprime}
    {d \rho^4 \over d{\Psi}}  = 2\: { ({\Psi}^3+2q^2 - \sqrt{6} q{\Psi})({\Psi}^3+2q^2 + \sqrt{6} q{\Psi})
    ({\Psi}^3-q^2)^4 \over {\Psi}^{17}}.
\end{equation}

Since $\Psi^3=q^2$ is a curvature singularity, we do not need to worry about
non-invertibility at that point. We need to focus on either $\Psi^3 > q^2$ or
$\Psi^3<q^2$. Since, asymptotically, $\Psi \simeq r^2$ is a large positive number, the
physical domain of the function $\Psi$ is
\begin{equation}\label{phy}
    q^{2/3} < \Psi < +\infty,
\end{equation}
and we explore invertibility of (\ref{rhoi}) on this domain.

All factors in (\ref{rhoprime}) are positive definite in the physical domain except for
the first one. We would then like to know if the solutions $\Psi_c$ to the equation $
{\Psi_c}^3+2q^2 - \sqrt{6} q{\Psi_c}=0$ lie in the physical range of the function $\Psi$
or not.  Let us first note this equation has positive solutions only if $q < q_c :=
\left( 2 / 3 \right)^{3/2}$. Thus, our first conclusion is that the relation (\ref{rhoi})
is invertible for $q>q_c$.

Recall now that the metric has two singularities, $\Psi^3=q^2$, and $\rho=0$. Let us call
$\Psi_0$ the particular value of $\Psi$ such that $\rho(\Psi_0) = 0$.  It turns out (this
is most easily done by a graphical analysis) that if $q < q_c$ the three numbers
$\Psi_c$, $\Psi_0$ and $q^{2/3}$ are ordered according to
\begin{equation}\label{order}
    q^{2/3} < \Psi_c < \Psi_0.
\end{equation}
This means that the non-invertible point $\Psi_c$ is beyond the origin $\rho=0$ and thus,
it does not affect the physical domain.

%---------------------------------------
\subsection{Horizon structure}

\begin{figure}[ht]
\centerline{\epsfig{file=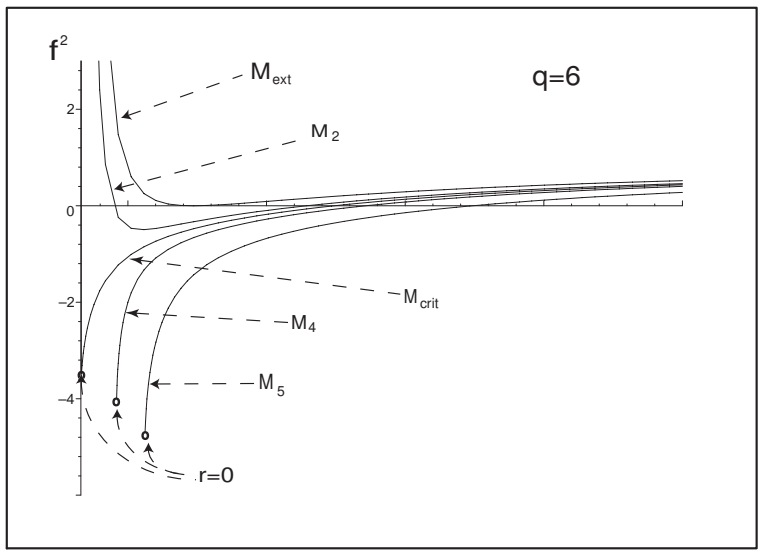,width=10cm,angle=0}} \caption{The function
 $f_1^2(\Psi)$ in the domain $q^{2/3} <\Psi<\infty$ for $q=6$ and five different values of $M$ }
 \label{Fig1}
\end{figure}

Given the form (\ref{metrici}) of the metric it is clear that horizons will arise
whenever the function $f_1^2$ vanishes. As mentioned before, the relation between $\rho$
and ${\Psi}$ is invertible and we can study $f_1^2$ as a function of ${\Psi}$. We write
here the explicit form of $f_1^2$ in terms of ${\Psi}$, the ADM mass $M$ and the charge
$q$,
\begin{eqnarray} \label{f1Psi}
  && f_1^2(\Psi) \ = \ 1  \ - \ {1 \over 2}\,{\frac {{q}^{2}}{{{\Psi}}^{2}}} \ - \ {\Psi} \ + \
   {{{\Psi}}^{6} \over \left( {{\Psi}}^{3}-{q }^{2}\right)^2}\times  \\
  &&\sqrt{{{\Psi}}^{2}-{1 \over 2}\,
         {\frac {{q}^{12}}{{{\Psi}}^{16}}}+{\frac {6{q}^{2}}
{{{\Psi}}^{2}}}+{\frac {24}{13}}{\frac {{q}^{10}}{{{\Psi}}^{13}}}- {9 \over 5}{\frac
{{q}^{8}}{{{\Psi}}^{10}}}-{\frac {48}{11}}{\frac {{q}^{8}}{{{\Psi}}^{11}}}-{ \frac
{8}{7}}{\frac {{q}^{6}}{{{\Psi}}^{7}}}+{6 \over 7}\,{\frac {{q}^{10}}{{{\Psi}}^{ 14}}} +
{\frac {9{q}^{6}}{{{\Psi}}^{8}}}+{\frac {3{q}^{4}}{{{\Psi}}^{4}}}-{ \frac
{48}{5}}\,{\frac {{q}^{4}}{{{\Psi}}^{5}}}-4M} \nonumber
\end{eqnarray}
In Fig. \ref{Fig1} we have plotted $f_1({\Psi})^2$ in the domain $q^{2/3} < \Psi <
\infty$, for $q=6$, and five different values of the mass $M$. (The picture is actually
generic for all values $q>q_c$.) Let us analyze each curve separately.

%FIG 1

%----------------------------
\subsubsection{Light black holes, $M<M_{crit}$, and extreme black holes}

The lightest case, corresponding to $M_{ext}=7.057...$, represents the extreme black
hole. It touches the horizontal line once, and its derivative is zero there too. The
masses $M_{ext}(q)$ are defined by the equations
\begin{equation}\label{extremality}
    f_1^2 =0 , \ \ \ \ \ {d f_1^2 \over d\Psi } =0.
\end{equation}
These equations can be solve numerically and we have found a {\it linear} relation
\begin{equation}\label{Mext}
    M_{ext}(q) \simeq -0.30914... + (1.2247...)\: q.
\end{equation}
(The linear approximation is better for charges $q > 2$.) This result is remarkable
because the equations (\ref{extremality}) defining $M_{ext}(q)$ form a extremely
non-linear system.  Note also that the slope of the curve approaches $\: 1.2247... =
\sqrt{3/2}$, which is precisely the value obtained by the asymptotic solution (valid for
$q/r^2 \ll 1$),
$$ f_1^2 \simeq 1 - {2M \over r^2} + {2
\over 3} {q^2 \over r^4} = \left( 1 -\sqrt{{2 \over 3}} {q \over r^2} \right)^2, \ \ \ \
\ \ \  M=\sqrt{{3 \over 2}}\: q$$ A word of caution is in order here. This analysis does
not imply that the function $M_{ext}(q)$ is exactly linear, for all values of $q$. We
only claim that the linear relation is a good approximation for that curve.

Let us now we move to the curve $M_2=7.8$. This  looks very much like a standard charged
black hole. $f_1^2$ intersects the horizontal line twice, and thus there are two
horizons.

The black holes discussed so far have one or two horizons, and $f_1^2$ diverges as one
approaches ${\Psi} \rightarrow q^{2/3}$; in these cases, the Kasner singularity at
$\psi^3=q^2$ is met {\it before} the origin $\rho=0$.

%-------------------
\subsubsection{The critical mass $M=M_{crit}$}

If we carry on making the black hole heavier, we reach the curve (for $q=6$)
$M_{crit}=8.113...$ where something new happens (we give a close expression for
$M_{crit}$ below). This curve intersects the horizontal line only once, and thus it has
only one horizon. Also, at the origin, $f_1^2$ has a finite value.

To have a better understanding of this case, consider the function $f_1^2$ displayed in
(\ref{f1Psi}).  For generic values of $M$ there is a explicit singularity at
${\Psi}^3=q^2$. However, if $M$ is fine-tuned such that the numerator (square root)
vanishes at that point, the pole is cancelled. In fact, one observes that the zero in the
square root is stronger than the zero in the denominator, and that whole term vanishes at
${\Psi}^3=q^2$.  The value of $f_1^2$ at that point is then,
\begin{equation}\label{q23}
    \left. f_1^2\right|_{{\Psi}^3=q^2} = 1 - {3 \over 2}\: q^{2/3},
\end{equation}
which is in fact finite. Since $M$ enters linearly in the square root, the value of $M$,
called $M_{crit}$, such that the square root vanishes at ${\Psi}^3=q^2$ can be calculated
directly,
\begin{eqnarray} \label{Mcrit}
  M_{crit} = {\frac {3^9}{8 \cdot 13!!}}\,{q}^{2/3} \left( 33\,{q}^{2/3}+26 \right),
\end{eqnarray}
($13!! \equiv 13*11*9\cdots 3*1$),

Finally, recall that the square root is nothing but $\rho^2$ (see Eq. (\ref{rhoi})). This
means that, by definition, at the critical mass $M_{crit}$, the origin $\rho=0$ and the
Kasner singularity $\Psi^3=q^2$ coincide.

%--------------------
\subsubsection{Heavy black holes:  $M>M_{crit}$}

Let us now increase the value of $M$ above $M_{crit}$. We find the curves $M_4=8.5$ and
$M_5=10$. These curves intersect the horizontal line only once. The associated black
holes then have only one horizon, despite being charged.

In this class of solutions (with $M>M_{crit}$) the origin $\rho=0$ is met {\it before}
the Kasner singularity. This is the reason that the curve stops before reaching
${\Psi}^3=q^2$. At $\rho=0$, $f_1^2$ has a finite value (just like the uncharged black
hole discussed in the introduction).

%---------------------
\subsubsection{The critical charge}

We have seen that the spectrum of black holes is separated into two branches, the heavy
branch with $M>M_{crit}$ and the light branch with $M<M_{crit}$. The interphase is
defined by the critical curve $M=M_{crit}$ displayed in (\ref{Mcrit}) which depends on
the charge $q$. We shall now see that there exists a particular value of $q$, namely,
\begin{equation}\label{qci}
    q_c = \left( {2 \over 3} \right)^{3/2},
\end{equation}
for which the light branch produces only naked singularities.

\begin{figure}[ht]
\centerline{\epsfig{file=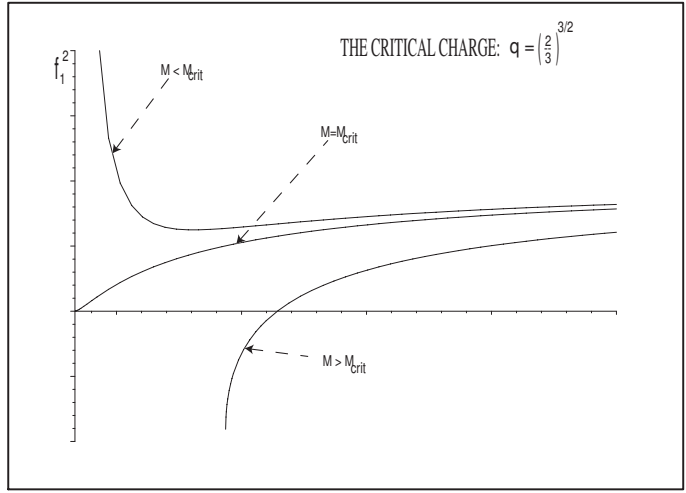,width=9cm,angle=0}} \caption{The function
$f_1^2$ as a function of ${\Psi}$ for $q=(2/3)^{3/2}$.} \label{Figqc}
\end{figure}

In fact, going back to Eq. (\ref{q23}) we note that for $q=q_c$, the value of $f_1^2$ at
$\Psi^3=q^2$ is zero. We have plot in Fig. \ref{Figqc} the function $f_1^2$ for $q=q_c$
and three different masses, $M <M_{crit}$, $M=M_{crit}$ and $M> M_{crit}$. We observe
that the curve $M < M_{crit}$ cannot intersect the horizontal line. The light branch thus
gives rise only to naked singularities.

For charges $q<q_c$ the critical curve is pulled upwards.  The light branch will carry on
producing naked singularities, while the heavy branch will have both, black holes and
naked singularities depending on $M$.

These results are summarized in  the ``phase diagram" displayed in the introduction, Fig.
\ref{Phases}, showing the various black holes types for all values of $M$ and $q$. The
most important aspect of that diagram, and of this paper, is the existence of two types
of black holes for charges $q>q_c$, which are continuously connected by varying the
parameters $M$ and $q$.

%--------------------------------------
\subsection{The Coulomb potential}

So far we have only analyzed the properties of the metric.  To compute the value of the
electrostatic potential we first go back to the redefinitions (\ref{redef}), and recall
that the functions $N_1,\psi_1$ and $\Psi_1$ are known in terms of $\Psi$, which is
algebraically related to the radial coordinate.  The full expression for the potential is
not very illuminating so we don't display it here. We only quote the result in the
asymptotic limit $r\rightarrow\infty$
\begin{equation}
    \phi(r) \simeq -3\sqrt{6}\: {q \over r^2} + {\cal O} \left( {1 \over r^4} \right)
\end{equation}
showing as stated above that the electromagnetic potential is asymptotically controlled
by Maxwell equations.

%------------
\acknowledgments

The author would like to thank Guido Tarrach for useful conversations, and  Kayll Lake
for introducing him into GRTensor II (http://grtensor.org/), which was very helpful in
various calculations. This work was partially supported by grant \# 1000744 (FONDECYT,
Chile), and from Ecos(France)-Conicyt(Chile) grant \# C01E05.


\begin{thebibliography}{10}

%\cite{Lovelock:1971yv}
\bibitem{Lovelock}
D.~Lovelock,
%``The Einstein Tensor And Its Generalizations,''
J.\ Math.\ Phys.\  {\bf 12} (1971) 498.
%%CITATION = JMAPA,12,498;%%

%\cite{Zwiebach:1985uq}
\bibitem{Zwiebach}
B.~Zwiebach,
%``Curvature Squared Terms And String Theories,''
Phys.\ Lett.\ B {\bf 156}, 315 (1985).
%%CITATION = PHLTA,B156,315;%%

\bibitem{Branes}  See, for example,
%\cite{Nojiri:2001ae}
S.~Nojiri, S.~D.~Odintsov and S.~Ogushi, Phys.\ Rev.\ D {\bf 65}, 023521 (2002)
[arXiv:hep-th/0108172];
%%CITATION = HEP-TH 0108172;%%
%\cite{Gregory:2003px}
J.~P.~Gregory and A.~Padilla, Class.\ Quant.\ Grav.\ {\bf 20}, 4221 (2003)
[arXiv:hep-th/0304250];
%%CITATION = HEP-TH 0304250;%%
%\cite{Padilla:2003qi}
A.~Padilla, Class.\ Quant.\ Grav.\ {\bf 20}, 3129 (2003) [arXiv:gr-qc/0303082];
%%CITATION = GR-QC 0303082;%%
%\cite{Gravanis:2002wy}
E.~Gravanis and S.~Willison, Phys.\ Lett.\ B {\bf 562}, 118 (2003)
[arXiv:hep-th/0209076];
%%CITATION = HEP-TH 0209076;%%
%\cite{Binetruy:2002ck}
P.~Binetruy, C.~Charmousis, S.~C.~Davis and J.~F.~Dufaux, Phys.\ Lett.\ B {\bf 544}, 183
(2002) [arXiv:hep-th/0206089];
%%CITATION = HEP-TH 0206089;%%
%\cite{Deruelle:2003ur}
N.~Deruelle and C.~Germani, arXiv:gr-qc/0306116.
%%CITATION = GR-QC 0306116;%%
%\cite{Deruelle:2003tz}
N.~Deruelle and M.~Sasaki, Prog.\ Theor.\ Phys.\ {\bf 110}, 441 (2003)
[arXiv:gr-qc/0306032];
%%CITATION = GR-QC 0306032;%%
%\cite{Low:2000pq}
I.~Low and A.~Zee, Nucl.\ Phys.\ B {\bf 585}, 395 (2000) [arXiv:hep-th/0004124].
%%CITATION = HEP-TH 0004124.%%

\bibitem{Boulware-D}
D.G.~Boulware and S.~Deser,
%``String Generated Gravity Models,''
Phys.\ Rev.\ Lett.\ {\bf 55}, 2656 (1985).

\bibitem{Wheeler} J.T. Wheeler, {\em Nucl.Phys.} {\bf B268}, 737
     (1986); {\bf B273}, 732 (1986).

\bibitem{Other}  For other aspects of Gauss-Bonnet black holes see,
%\cite{Neupane:2002bf}
I.~P.~Neupane, Phys.\ Rev.\ D {\bf 67}, 061501 (2003) [arXiv:hep-th/0212092];
%%CITATION = HEP-TH 0212092;%%
%\cite{Cho:2002hq}
Y.~M.~Cho and I.~P.~Neupane, Phys.\ Rev.\ D {\bf 66}, 024044 (2002)
[arXiv:hep-th/0202140];
%%CITATION = HEP-TH 0202140;%%
%\cite{Cvetic:2001bk}
M.~Cvetic, S.~Nojiri and S.~D.~Odintsov, Nucl.\ Phys.\ B {\bf 628}, 295 (2002)
[arXiv:hep-th/0112045];
%%CITATION = HEP-TH 0112045;%%
S.~Nojiri and S.~D.~Odintsov, Phys.\ Rev.\ D {\bf 66}, 044012 (2002)
[arXiv:hep-th/0204112].
%%CITATION = HEP-TH 0204112;%%

\bibitem{BTZ2} M. Ba\~nados, C. Teitelboim and J.Zanelli,
               Phys. Rev. {\bf D49}, 975 (1994).

%\cite{Deser:tn}
\bibitem{Deser-JtH}
S.~Deser, R.~Jackiw and G.~'t Hooft, Annals Phys.\  {\bf 152}, 220 (1984).
%%CITATION = APNYA,152,220;%%

\bibitem{Chamseddine} A.H. Chamseddine,
Nucl. Phys. {\bf B346}, 213 (1990).

%\cite{Banados:2001hm}
\bibitem{B}
M.~Ba\~nados,
%``Charged solutions in 5d Chern-Simons supergravity,''
Phys.\ Rev.\ D {\bf 65}, 044014 (2002) [arXiv:hep-th/0109031].
%%CITATION = HEP-TH 0109031;%%

\bibitem{Achucarro} A. Ach\'ucarro, P.K. Townsend,  Phys. Lett. {\bf B180}, 89 (1986). E. Witten,  Nucl. Phys. {\bf B 311}, 4
(1988).

\bibitem{CS} See, for example, G.~V.~Dunne and C.~A.~Trugenberger,
Annals Phys.\  {\bf 204}, 281 (1990); V.P. Nair, J. Schiff, Phys. Lett. {\b B246}, 423,
(1990);  Nucl. Phys. {\bf B371}, 329 (1992); M. Ba\~nados, L.J. Garay, M. Henneaux, Phys.
Rev. {\bf D53}, R593 (1996); Nucl. Phys. {\bf B476}, 611 (1996); R. Troncoso and J.
Zanelli, Phys.Rev. {\bf D58}, R101703 (1998); J.~Gegenberg and G.~Kunstatter, Phys.\
Lett.\  {\bf B478}, 327 (2000); P.~Horava, Phys.\ Rev.\ {\bf D59}, 046004 (1999)
[hep-th/9712130].

\bibitem{BTZ} M. Ba\~nados, C. Teitelboim and J.Zanelli,
              Phys. Rev. Lett. {\bf 69}, 1849 (1992).
              Ba\~nados, M. Henneaux, C. Teitelboim and
              J.Zanelli,  Phys. Rev. {\bf D48}, 1506 (1993).

%\cite{Alfaro:2003hk}
\bibitem{Alfaro-B}
J.~Alfaro and M.~Banados, to appear in Phys.Rev.D, arXiv:hep-th/0307249.
%%CITATION = HEP-TH 0307249;%%

\end{thebibliography}
\end{document}